\documentclass[aps,prl,twocolumn,groupedaddress,showpacs]{revtex4}
\usepackage{graphicx}

\begin{document}

\title{Energy barriers in spin glasses}


\author{Barbara Drossel}
\affiliation{Institut f\"ur Festk\"orperphysik,  TU Darmstadt,
Hochschulstra\ss e 6, 64289 Darmstadt, Germany }
\author{M.A. Moore}
\affiliation{Department of Physics, University of Manchester, 
Manchester M13 9PL, U.K.}

\date{\today}

\begin{abstract}
For an Ising spin glass on a hierarchical lattice, we show that the
energy barrier to be overcome during the flip of a domain of size L
scales as $L^{d-1}$ for all dimensions $d$. We do this by investigating 
appropriate lower bounds to the barrier energy, which can be evaluated
using an algorithm that remains fast for large system sizes
  and dimensions. The asymptotic limit $d\to \infty$ is evaluated
  analytically.
\end{abstract}


\maketitle

Energy barriers determine the dynamics of glassy systems that have a
complex energy landscape with many metastable states. Typically, the
fluctuations between free energy minima in these systems (either
between different realizations or in the same random system) scale
with observation size $L$ as $L^\theta$. It is generally assumed that
the free energy barriers encountered in moving from one minimum to
another scale with observation size as $L^\psi$. However, there exist
so far few numerical studies of the value of this exponent in
spin glasses, probably because of the difficulty of the problem
(in contrast to the many studies of $\theta$). For directed
polymers in random systems, the identity $\psi=\theta$ was demonstrated some
time ago \cite{ourbarrierpapers}. For Ising spin glasses, Fisher and
Huse derived within the droplet picture the double inequality $\theta
\le \psi \le d-1$ \cite{fisherhuse} for dimension $d$. The lower limit is due to the
fact that a domain wall has to be introduced into the system if all
its spins are to be flipped. However, the minimum domain wall energy
scales as $L^{\theta}$. The upper limit is obtained by moving a
straight domain wall through the system. Since such a domain wall
breaks $L^{d-1}$ bonds, its energy cannot be larger than
 $\sim L^{d-1}$.  Experiments on two- \cite{schins93} and
three-dimensional \cite{jonss02} spin glasses as well as a numerical
studies in two dimensions \cite{tom91} point towards a value of
$\psi$ close to or identical to $d-1$. On the other hand, an equality
$\psi=\theta$ is sometimes tacitly assumed, as for instance in a
recent publication on spin glass dynamics on the hierarchical lattice
\cite{sas03}, where the probability for a spin flip on the length
scale $L$ is chosen to be a function of the effective coupling
strength on this scale, which increases as $L^\theta$.

In this paper, we consider the Ising spin glass on an hierarchical
lattice and show in fact that $\psi = d-1$ in all dimensions. The Hamiltonian of
the system is given by 
\begin{equation}
H = -\sum_{\langle ij\rangle} J_{ij}S_i S_j\, ,
\end{equation}
where $\langle ij\rangle$ indicates the sum over all nearest-neighbor
pairs, and the spins assume the values $\pm 1$. We will mainly
consider a Gaussian distribution of couplings with zero mean and unit
width. The hierarchical lattice gives for the exponent 
$\theta$ in dimension $d=2$ the value $-0.27$ \cite{BM84}
which compares well with the value $-.29$ from recent numerical studies on
the square lattice \cite{HM} while in three dimensions the hierarchical
lattice gives  $\theta \approx0.25$ \cite{BM84} while on a simple cubic lattice its value is
close to $0.20$ \cite{AKH99}. Thus the hierarchical lattice provides reasonably good 
estimates for the value of the exponent $\theta$ (at least for
low-dimensional systems) and it is our hope that it is equally
useful for determining the value of $\psi$.

The problem of
finding energy barriers in glassy systems numerically is usually
NP-complete \cite{mid99}. We will therefore not attempt to calculate
the barrier exactly, but we will rather place bounds on it. Since the
upper bound $\psi=d-1$ is already known due to the above-cited
argument by Fisher and Huse, we will show in the following that there
exists a lower bound to the barrier energy that increases with system
size as $L^{d-1}$.

\begin{figure}
\includegraphics*[width=8.5cm]{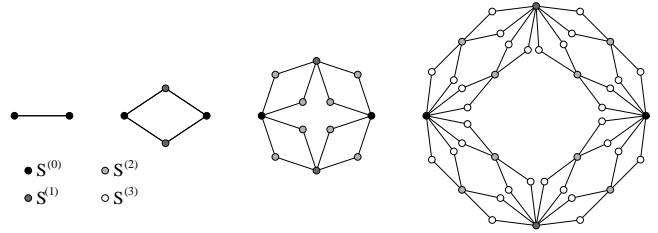}
\caption{Iterative construction of the hierarchical lattice. At each
  step, each bond is replaced with $2^{d-1}$ pairs of bonds, with a
  new lattice site in between. Different grey shades indicate sites added
  at different iteration steps.
\label{fig1} 
}
\end{figure}

A hierarchical lattice is constructed by starting with one bond
connecting two sites. This bond is replaced with a unit consisting of
$2^{d-1}$ pairs of bonds, with a new site between each pair. Each of
the $2^d$ bonds is again replaced with a unit of
$2^{d-1}$ pairs of bonds, etc., leading to a lattice with $2^{Id}$
bond after $I$ iterations. In Fig.~\ref{fig1} the first three
steps of this process are illustrated for $d=2$. 
Evaluating a thermodynamic quantity on the hierarchical lattice with
$I$ levels is equivalent to evaluating it on a $d$-dimensional
hypercubic lattice of linear size $2^I$ using the
Migdal-Kadanoff-approximation. Recently, also the dynamics of spin glasses have
been studied on hierarchical lattices \cite{ric00,sas03,sch03}, although there is no
simple relation to the dynamics on hypercubic lattices. In
\cite{sas03,sch03}, approximations based on renormalization ideas were
made.

\begin{figure}
\includegraphics*[width=8.5cm]{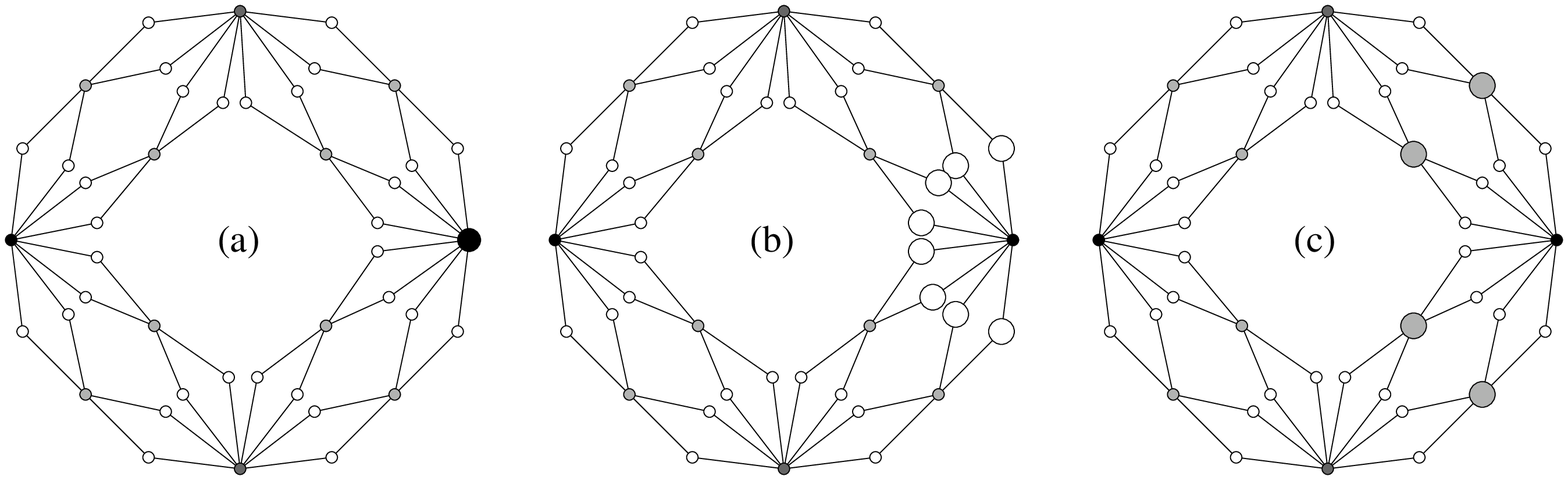}
\caption{(a) The right-hand corner site drawn large, (b) its
  nearest-neighbors drawn large, (c) its next-nearest neighbors drawn large.
\label{fig2} 
}
\end{figure}

We focus on the energy barrier that has to be overcome when moving
from the ground state to the lowest-energy configuration with a domain
wall. This domain-wall configuration is obtained by flipping one of
the two level-0 spins and by determining the new ground state with
this fixed new configuration of the level-0 spins. We allow only
single-spin flips when moving from the initial to the final state. At
zero temperature (the situation we are considering here), the
free-energy barrier is identical to the energy barrier.  As indicated
above, our goal is to show that there exists a lower bound that scales
as $L^{d-1}$. For this purpose, we consider the neighborhood of the
right-hand level-0 spin, the ``corner spin'' (see Fig.~\ref{fig2}).
We focus our attention on its $L^{d-1}$ level-$I$ nearest neighbors
and its $(L/2)^{d-1}$ level-$(I-1)$ next-nearest neighbors. At the
moment where the corner spin is flipped, the next-nearest neighbors
are in a configuration $\mathcal{C}_{nnn}$. We do not know the
configuration of these spins which is associated with the true
barrier, but we know that the optimum spin-flip sequence which passes
through the true barrier state must have one of the possible
configurations $\mathcal{C}_{nnn}$ of the next-nearest neighbor spins
at the moment where the corner spin is flipped. Therefore we will
later minimize our lower bound with respect to $\mathcal{C}_{nnn}$. We
start from the configuration of lowest energy that can be obtained
with a given configuration $\mathcal{C}_{nnn}$ of the next nearest
neighbors of the right-hand corner spin and with the two corner spins
fixed at their initial configuration. Clearly, the energy of this
state is at least as high as the energy of the initial configuration.
We next calculate the minimum energy barrier that has to be overcome
when the right-hand corner spin is flipped with the configuration of
the next-nearest neighbors remaining fixed. This minimum energy
barrier is obtained by first flipping a suitable selection of the
nearest-neighbor spins of the corner spin, before the corner spin
itself is flipped. The barrier state is the one immediately before or
immediately after the corner spin is flipped, and it is reached from
our initial state by spin flips each of which increases the energy.
Minimizing the energy barrier (i.e. the energy difference between the
barrier state and our initial state) with respect to
$\mathcal{C}_{nnn}$ gives a lower bound to the true barrier. This is
because the energy of our initial configuration is at least as high as
that of the true initial configuration and since the energy of our
barrier state cannot be larger than that of the true barrier state.

In the following, we determine this lower bound to the barrier. The
right-hand corner spin is connected to each of its next-nearest
neighbors via $2^{d-1}$ intermediate (level-$I$) spins, each of which
initially assume the orientation that has the lower energy.  Our task
now consists in finding those intermediate spins that have to be
flipped before the corner spin is flipped, such that the barrier
energy is as low as possible. Intermediate spins, for which the
absolute value of the coupling to its left neighbor is stronger than
the absolute value of the coupling with the corner spin, must not be
flipped. This is because for a given configuration of its two
neighboring spins, the energy is always lower when the intermediate
spin has the orientation that satisfies the left-hand bond and
possibly frustrates the right-hand bond. Since the values of the
couplings are assigned at random, only about half of the intermediate
spins are candidates for being flipped before the corner spin.

\begin{figure}
\includegraphics*[width=7cm]{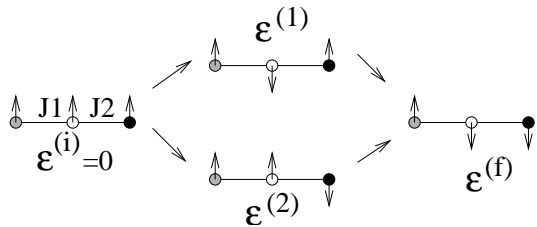}
\caption{A subunit of three spins and the energies associated with the
  different spin configurations mentioned in the text. The figure is
  drawn for the case that the right-hand bond is stronger than the
  left-hand bond ($|J2| > |J1|$).
\label{fig3} 
}
\end{figure}

Evaluating all possible combinations of these intermediate spins that
might be flipped before the corner spin costs a computer time that
increases exponentially with the number of these spins. We will
therefore later make an approximation that underestimates the
above-defined lower bound to the barrier. In order to define the
quantities we need, let us first consider a subunit of three spins
(see Fig.~\ref{fig3}): The corner spin on the right, one of its
next-nearest neighbors to the left of it, and the intermediate spin
sitting between the two and connected to both of them. The left-hand
spin is fixed. The corner spin is first in its initial
configuration. The intermediate spin has the configuration that
minimizes the energy. This initial energy is our reference energy, and
we set it to zero. Now let $\epsilon^{(1)}$ be the energy of the
three-spin unit when the intermediate spin is flipped, and let
$\epsilon^{(2)}$ be the energy of the three-spin unit when the
right-hand spin (the corner spin) is flipped without first flipping
the intermediate spin, and let $\epsilon^{(f)}$ be the energy of the
three-spin unit when the right-hand spin is flipped and the
intermediate spin is adjusted such that it minimizes the energy. If
the left-hand bond is stronger than the right-hand bond, we have
$\epsilon^{(2)}=\epsilon^{(f)}$, otherwise we have
$\epsilon^{(2)}>\epsilon^{(f)}$.

Now let $n$ count all those three-spin units for which the
intermediate spin is being flipped before the corner spin, and let $m$
count all those  three-spin units for which the
intermediate spin is not flipped before the corner spin. 
The energy of the system just before the corner spin is flipped is
$$\sum_n \epsilon_n^{(1)} \equiv E_B^{(1)}$$ 
and the energy of the system right after
the corner spin is flipped is 
$$\sum_m \epsilon_m^{(2)}+ \sum_n \epsilon_n^{(f)} \equiv E_B^{(2)}$$

The lower
bound we are looking for is then
\begin{eqnarray}
E_B&=&
Min_{C_{nnn}} Min_{n,m} Max\left[E_B^{(1)},E_B^{(2)} \right]\nonumber\\
&\ge& Min_{C_{nnn}}Min_{n,m}\left[\frac 1 2 (E_B^{(1)}+ E_B^{(2)}) \right]\nonumber\\
&=&Min_{C_{nnn}} Min_{n,m}\left[\frac 1 2 \left(\sum_n(\epsilon_n^{(1)}+ \epsilon_n^{(f)} ) + \sum_m \epsilon_m^{(2)}
\right) \right]\nonumber\\
&=&Min_{C_{nnn}} \frac 1 2\sum_i Min\left[\epsilon_i^{(1)}+
  \epsilon_i^{(f)},\epsilon_i^{(2)} \right] \equiv S\, .\label{mainresult}
\end{eqnarray}
\begin{figure}
\includegraphics*[width=7cm]{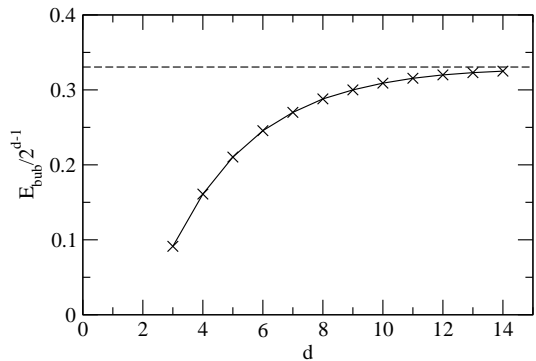}
\caption{$E_{bub}/2^{d-1}$ as function of dimension $d$. The dashed
  horizontal line indicates the analytically determined asymptotic
  value 0.3305 which is approached in the limit $d\to\infty$.
\label{fig4} 
}
\end{figure}

Since $S$ underestimates $E_B$, it is also a lower bound to the
barrier, and we focus on it in the following.  We now perform the
minimization with respect to the configuration of the next-nearest
neighbor spins, $\mathcal{C}_{nnn}$. For each ``bubble'' consisting of
the corner spin, a next-nearest neighbor and $2^{d-1}$ intermediate
spins, the contribution to the above sum is minimized if the
next-nearest neighbor is in the configuration that leads to the higher
initial bubble energy. We are interested in the average of the above
sum over many different systems. Since the contributions of the
bubbles to this average are additive, it is sufficient to take the
average over one bubble, and our lower bond $S$ is then for large $L$
simply the number of bubbles, $(L/2)^{d-1}$, times this average. 
The lower bound to the barrier is therefore for
large $L$ given by
\begin{equation}
S=(L/2)^{d-1}\times E_{bub} \sim L^{d-1} \,
\end {equation}
i.e., it scales as $L^{d-1}$.  For a Gaussian distribution of unit
width of the couplings, the bubble average has in three dimensions the
value $E_{bub}=0.364$, i.e. it is positive. It increases with
increasing dimension and approaches for large dimensions the asymptotic
value $E_{bub}=0.3305 \times 2^{d-1}$ (see Fig.~\ref{fig4}).  

For the limit $d\to \infty$ we can prove analytically that $E_{bub}$
is positive. For each three-spin unit, we call the weaker coupling
$J1$, and the stronger coupling $J2$. In the limit $d\to \infty$, one
quarter of all three-spin units are initially not frustrated and have
the stronger bond on the left-hand side. They make together a
contribution $2^{d-3}\langle|J1|\rangle$ to $E_{bub}$. One quarter of
all three-spin units are initially not frustrated and have the weaker
bond on the left-hand side. They make together a contribution
$2^{d-3}\langle|J2|\rangle$ to $E_{bub}$. One quarter of all
three-spin units are initially frustrated in the right-hand bond and
make together a contribution $-2^{d-3}\langle|J1|\rangle$ to
$E_{bub}$. The last quarter of all three-spin units are initially
frustrated in the left-hand bond and make together a contribution
$2^{d-3}\langle|J2|-2|J1|\rangle$ to $E_{bub}$. Together this gives
$E_{bub}=2^{d-2}\langle|J2|-|J1|\rangle$, which is positive since
$|J2|>|J1|$. For a Gaussian bond distribution of unit width, $|J1|=
2\frac{\sqrt{2}-1}{\sqrt{\pi}}=0.467$ and $|J2|=
\frac{2}{\sqrt{\pi}}= 1.128$, leading to $E_{bub}=0.3305\times
2^{d-1}$, in agreement with our asymptotic limit in Fig.~\ref{fig4}.

Our lower bound $S$ obtained by calculation (\ref{mainresult}) is
negative for $d=2$ and is therefore useless in this case. The reason
is that the value of $\epsilon^{(2)}$ is more often negative in $d=2$
than in higher dimensions (because we start from the higher initial
bubble energy), leading to a large difference between $E_B^{(1)}$ and
$ E_B^{(2)}$ and making the inequality in the second line of
(\ref{mainresult}) worse than in higher dimensions. In order to obtain
a positive  $E_{bub}$ also in $d=2$, we replace
in the second line of (\ref{mainresult}) the term $\frac 1 2
(E_B^{(1)} +E_B^{(2)})$ with the more general expression $a E_B^{(1)}
+ (1-a)E_B^{(2)}$, which is valid for all $ 0<a<1 $. We then obtain
$$ S=Min_{C_{nnn}} \sum_i Min\left[a\epsilon_i^{(1)}+ (1-a)
 \epsilon_i^{(f)},(1-a)\epsilon_i^{(2)} \right]\, .$$ Choosing $a>0.5$
 places more weight on the larger energy $E_B^{(1)}$ and makes
 $E_{bub}$ indeed positive.  Already for $a=0.51$ we obtain a positive
 $E_{bub}/2\simeq 0.0039$, and for $a=0.74$, it is as large as 0.12.
 In $d=1$, the barrier energy is twice the absolute value of the
 largest bond, which increases as $\sqrt{\ln L}$ for a Gaussian
 distribution of bonds. We therefore obtain (apart from logarithmic
 corrections in $d=1$) the general result that the barrier scales on
 the hierarchical lattice for all $d$ as $L^{d-1}$, implying
 $\psi=d-1$. 

\begin{figure}
\includegraphics*[width=7cm]{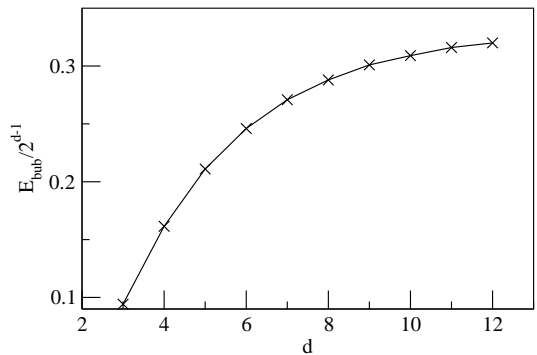}
\caption{The figure shows a plot of $E_{bub}/2^{d-1}$ for a $J=\pm
   2^{(1-d)/2}$ bond distribution after one decimation step as
   function of dimension $d$.  \label{fig5}  }
\end{figure}

 In the following, we argue that our results can be
 generalized to other bond distributions. For some distributions, our
 expression for $E_{bub}$ can become negative. This happens for
 instance for the $\pm J$ model, where the final state of the bubble
 can be reached from the initial state without first increasing the
 energy ($E_{bub}=-1.5$ for $d=3$). In this case we modify our
 calculation in the following way: we first perform a sufficient
 amount of decimation steps on the hierarchical lattice, until the
 distribution of the renormalized bonds is such that our expression
 $S$ (Eq.~(\ref{mainresult})) for the lower bound becomes positive. It
 must eventually become positive since the asymptotic bond
 distribution (if rescaled to unit width) is for all distributions
 with finite moments the same and is very close to a Gaussian
 distribution.  Then our argument can be repeated on the
 coarse-grained lattice, giving a new lower bound to the barrier
 energy for the original lattice. This is because the energy of the
 system with a given configuration of those spins that survived the
 renormalization procedure can never be smaller than that of the
 renormalized system. It turned out that performing one decimation
 step is sufficient for the $\pm J$ model even in $d=3$ in order to
 obtain a positive value of $E_{bub}$.  Fig.~\ref{fig5} shows
 $E_{bub}/2^{d-1}$ as function of $d$ for the distribution of bonds
 that is obtained after one decimation step for bond values $J=\pm
 2^{(1-d)/2} $. (The values are chose such that the width of the
 distribution after one decimation step is 1.) The data are almost
 indistinguishable from those of Fig.~\ref{fig4}. This is to be
 expected in higher dimensions, since the bond distribution after one
 decimation step is a binomial distribution, which approaches a
 Gaussian in high dimensions.
 We therefore obtain even for the $\pm J$ system a lower bound to the
barrier that scales as $L^{d-1}$. However, we can expect strong
finite-size effects for small system sizes.

To conclude, we have shown that the energy barrier that has to be
overcome when introducing a domain wall into an Ising spin glass on a
hierarchical lattice scales in all dimensions as $L^{d-1}$, for all
bond distributions with finite moments. It remains to be seen if these
results can be generalized to conventional lattices. However, we find it 
encouraging that the experimental data for $\psi$ seem strongly to
favor a value close to $d-1$ rather than the \lq\lq rival\rq\rq 
 value $\theta$. Since experimental results on spin glasses are  not always probing 
droplets on the length scales at which droplet scaling ideas can be 
expected to apply without the use of corrections to scaling \cite{HM}, it
is unrealistic to expect perfect agreement for the value of the exponent
$\psi$. A further complication is that experimental and numerical data are
often studied at temperatures quite close to the transition temperature $T_c$,
when crossover to critical behavior will also complicate the analysis 
\cite{MBD}.

\acknowledgements{This work was supported in part by ESF's SPHINX
programme.}

 \end{document}